\documentclass[reprint,amsmath,amssymb,aps,pre]{revtex4-2}
\usepackage{graphicx}
\usepackage{color}
\usepackage{amsmath}
\usepackage{amssymb}
\usepackage{amsthm}
\usepackage{extarrows}
\usepackage{tabularx}
\usepackage{booktabs}
\usepackage{float}
\usepackage[dvipdfm,colorlinks,urlcolor=blue,linkcolor=blue,anchorcolor=blue,citecolor=blue]{hyperref}
\usepackage{lineno}

\begin{document}

\title{The Structure of the Route to the Period-three Orbit in the Collatz Map}

\author{Weicheng Fu$^{1,2,3}$}
\email{fuweicheng@tsnu.edu.cn}
\author{Yisen Wang$^{3}$}
\email{wys@lzu.edu.cn}

\affiliation{
$^1$ Department of Physics, Tianshui Normal University, Tianshui 741000, Gansu, China\\
$^2$ Key Laboratory of Atomic and Molecular Physics $\&$ Functional Material of Gansu Province, College of Physics and Electronic Engineering, Northwest Normal University, Lanzhou 730070, China\\
$^3$ Lanzhou Center for Theoretical Physics, Key Laboratory of Theoretical Physics of Gansu Province, and Key Laboratory of Quantum Theory and Applications of MoE, Lanzhou University, Lanzhou, Gansu 730000, China\\
}

\date{\today }

\begin{abstract}
This study analyzes the Collatz map through nonlinear dynamics. By embedding integers in Sharkovsky's ordering, we show that odd initial values suffice for full dynamical characterization. We introduce ``direction phases'' to partition iterations into upward and downward phases, and derive a recursive function family parameterized by upward phase counts. Consequently, a logarithmic scaling law between iteration steps and initial values is revealed, demonstrating finite-time convergence to the period-three orbit. Moreover, we establish the equivalence of the Collatz map to a binary shift map, whose ergodicity guarantees universal convergence to attractors, providing additional support for convergence. Furthermore, we identify that basins of attraction follow power-law distributions and find that odd numbers classified by upward phases follow Gamma statistics. These results offer valuable insights into the dynamics of discrete systems and their connections to number theory.
\end{abstract}

\maketitle

\section{Introduction}

Dynamic systems exhibiting behaviors from periodic oscillations to chaotic phenomena are ubiquitous in nature and engineering \cite{RevModPhys.53.655,PhysRevLett.49.1801,PhysRevX.13.021018,PhysRevE.110.014201,PhysRevResearch.6.033128,PhysRevE.110.024215}. Nonlinear dynamics provides a systematic framework to analyze and predict the evolution of such complex systems \cite{fuchs2014nonlinear}. Discrete dynamical systems, governed by difference equations and characterized by iterative state transitions, serve as effective models for phenomena including population dynamics, digital signal processing, and economic cycles \cite{Abraham1997}. These systems exhibit diverse behaviors such as periodicity, chaos, bifurcations, and fractal patterns \cite{peitgen2004chaos,YAO2019100027,ROUAH2024100104}, with their theoretical significance and practical relevance establishing them as a central focus in nonlinear dynamics research \cite{Feigenbaum1978,CHIRIKOV1979263,ott2002chaos,RevModPhys.61.981,driebe1999fully,9321487,Oliveira2024,10813599}.

In nonlinear dynamics, Sharkovsky's theorem \cite{sharkowskii1964co, vstefan1977theorem} and Li-Yorke's theorem \cite{li1975period} are foundational results. Sharkovsky's theorem establishes the coexistence of periodic solutions in one-dimensional discrete maps, known as Sharkovsky's ordering \cite{kloeden1979sharkovsky}, demonstrating that specific periodic solutions imply the existence of others. Notably, a period-three solution guarantees solutions for all integer periods. Li-Yorke's theorem, encapsulated by ``Period three implies chaos'' \cite{li1975period}, both recontextualizes Sharkovsky's insight and rigorously defines chaos. It shows chaotic systems exhibit extreme sensitivity to initial conditions: infinitesimal differences in initial states diverge exponentially, precluding long-term predictability \cite{2007coestreicher, lorenz1963deterministic}. This discovery of deterministic chaos overturns classical assumptions that certainty ensures predictability. While Newtonian mechanics struggles with nonlinear systems, chaos theory redefines our understanding of complexity \cite{prigogine1997end, zaslavsky2005hamiltonian, gleick2008chaos}. Importantly, the ``Period three implies chaos'' principle applies exclusively to continuous interval self-maps and does not generalize to discrete maps.

In this study, we analyze the Collatz map as a discrete dynamical system, employing methods from nonlinear dynamics to investigate its statistical behavior \cite{Sinai2003}. The Collatz map also known as the Collatz conjecture (or $3X+1$ problem), renowned for its simple formulation \cite{wirsching1998dynamical} and unresolved status \cite{lagarias20113x1problemannotatedbibliography,lagarias20123x1problemannotatedbibliography,lagarias20213x1problemoverview,tao2022orbitscollatzmapattain,hu2021analysis,wang2022proof}, posits that iterating the transformation
\begin{equation}\label{eq-model}
X = \begin{cases}
3X + 1 & \text{if } X \text{ is odd}, \\
X/2 & \text{if } X \text{ is even},
\end{cases}
\end{equation}
will eventually reduce any positive integer $X$ to the cycle $\{4, 2, 1\}$.

Significant progress has been made in characterizing the probabilistic behavior of Collatz orbits. Foundational work by Crandall \cite{1978Crandall} established constraints linking the map to Diophantine equations, while subsequent studies like Krasikov’s statistical approximations \cite{Krasikov1989} advanced a probabilistic framework. Recent advances by Tao \cite{tao2022orbitscollatzmapattain} leverage random walks and skew distributions on cyclic groups to demonstrate convergence for ``almost all'' integers (in logarithmic density), offering strong heuristic support for the conjecture.

The Collatz conjecture hinges on two requirements: (i) proving \emph{uniqueness} (i.e., that $\{4,2,1\}$ is the sole periodic cycle), and (ii) demonstrating \emph{convergence} (i.e., finite iterations terminating at this cycle). Fulfilling both would resolve the conjecture, yet despite computational verification for astronomically large integers, a universal proof remains beyond reach.

In this work, we employ Sharkovsky's ordering to represent integers and prove that analyzing Collatz map iterations necessitates focusing solely on odd initial values. By introducing ``direction phases''---classifying iterations into upward and downward phases---we derive a recursive function family parameterized by upward-phase counts. This reveals a logarithmic relationship between iteration steps and initial values, demonstrating finite-time convergence to the period-three orbit $\{4,2,1\}$. Furthermore, we establish equivalence between the Collatz map and a binary shift map. The ergodicity of the shift map \cite{book2011Ergodic} ensures inevitable evolution toward attractors, reinforcing convergence. Numerical analysis shows that attraction basins obey power-law distributions, while odd numbers grouped by upward phases approximately follow Gamma distributions. However, directly proving the cycle's uniqueness remains unresolved, as discussed later. Our findings bridge discrete dynamics and number theory.

The paper is structured as follows: Section \ref{sec:2} defines the Collatz map and introduces the concept of ``direction phases'', Section \ref{sec:3} develops the theoretical framework, Section \ref{sec:4} validates results numerically, and Section \ref{sec:5} concludes with open challenges.

\section{The Collatz map and definition of direction phases}\label{sec:2}

For brevity, the map (\ref{eq-model}) is abbreviated as
\begin{equation}\label{eq-model-m}
  X_n = M(X_{n-1}) = M^n(X_0),
\end{equation}
where \( n, X_n \in \mathbb{Z}^+ \). The Collatz conjecture asserts
\begin{equation}\label{eq-C-conjuecture}
\lim_{n \to \infty} M^n(X_0) = \{4,2,1\} \quad \text{for } \forall X_0 \in \mathbb{Z}^+,
\end{equation}
i.e., all trajectories eventually enter the period-three orbit $\{4,2,1\}$.

\begin{figure}[t]
  \centering
  \includegraphics[width=1\columnwidth]{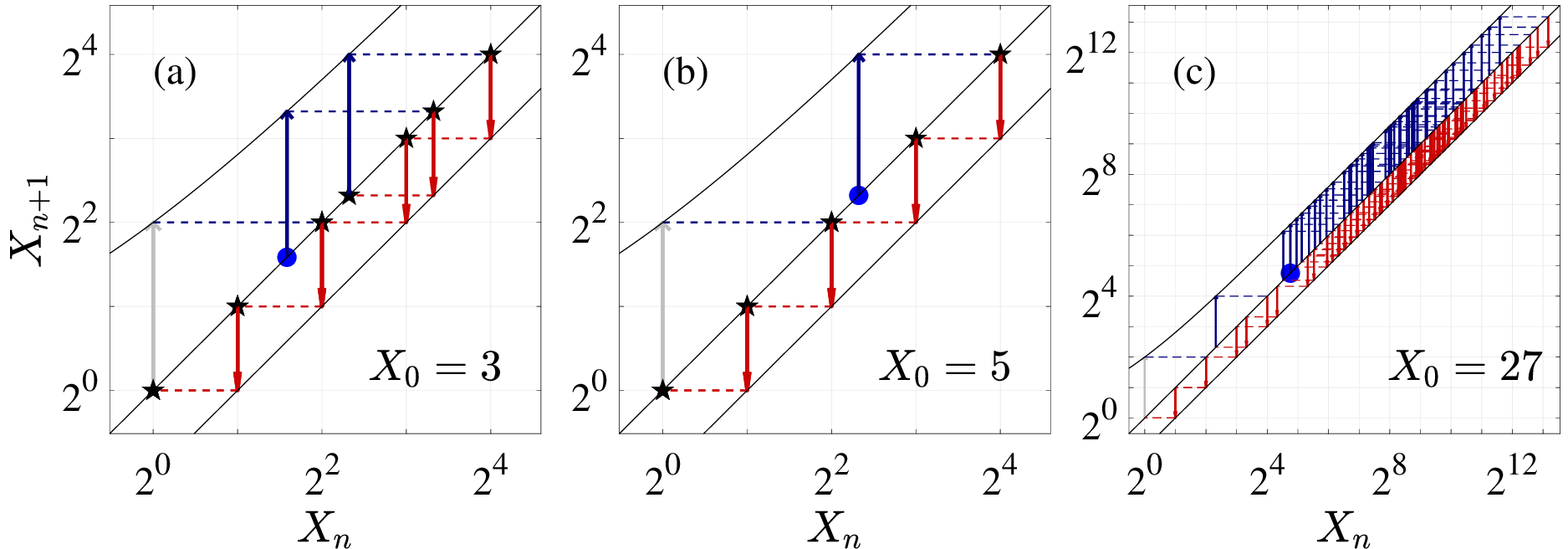}
  \caption{A cobweb plot for orbits of different $X_0$.}\label{fig-cobweb}
\end{figure}

To formalize the dynamics, we introduce ``direction phases'' \cite{Wang2000}:
\begin{equation}\label{eq-phase}
\begin{cases}
P_{\uparrow}(n) = 1, &\text{if} ~X_{n+1} > X_n\quad \text{(up phase)}; \\
P_{\downarrow}(n)= -1, &\text{if}~ X_{n+1} < X_n\quad \text{(down phase)}.
\end{cases}
\end{equation}
Let $N_{\uparrow}$ and $N_{\downarrow}$ denote the number of up and down phases \emph{before} entering $\{1,4,2\}$, respectively. For example, Figs.~\ref{fig-cobweb}(a)-(c) show trajectories for $X_0 = 3, 5, 27$, yielding $(N_{\uparrow}, N_{\downarrow}) = (2,5), (1,4), (41,70)$, respectively.

Notably, $N_{\uparrow}$ and $N_{\downarrow}$ correspond to the counts of odd and even terms in the sequence from $X_0 \neq 1$ to 2. The total iterations satisfy
\begin{equation}\label{eq-Nsum}
N = N_{\uparrow} + N_{\downarrow}, \quad \text{with } M^N(X_0) = 1.
\end{equation}

The Collatz map generalizes to all integers via
\begin{equation}\label{eq-g-C-map}
X_{n+1} =
\begin{cases}
3X_n + \text{sgn}(X_n) & \text{(odd } X_n\text{)},\\
{X_n}/{2} & \text{(even } X_n\text{)},
\end{cases}
\end{equation}
where \( X_n \in \mathbb{Z} \). The extended conjecture becomes
\begin{equation}\label{eq-g-C-map-end}
\lim_{n \to \infty} M^n(X_0) =
\begin{cases}
\{-1,-4,-2\} & X_0 \in \mathbb{Z}^-, \\
\{0\} & X_0 = 0, \\
\{4,2,1\} & X_0 \in \mathbb{Z}^+.
\end{cases}
\end{equation}

\begin{figure}[tbp]
  \centering
  \includegraphics[width=1\columnwidth]{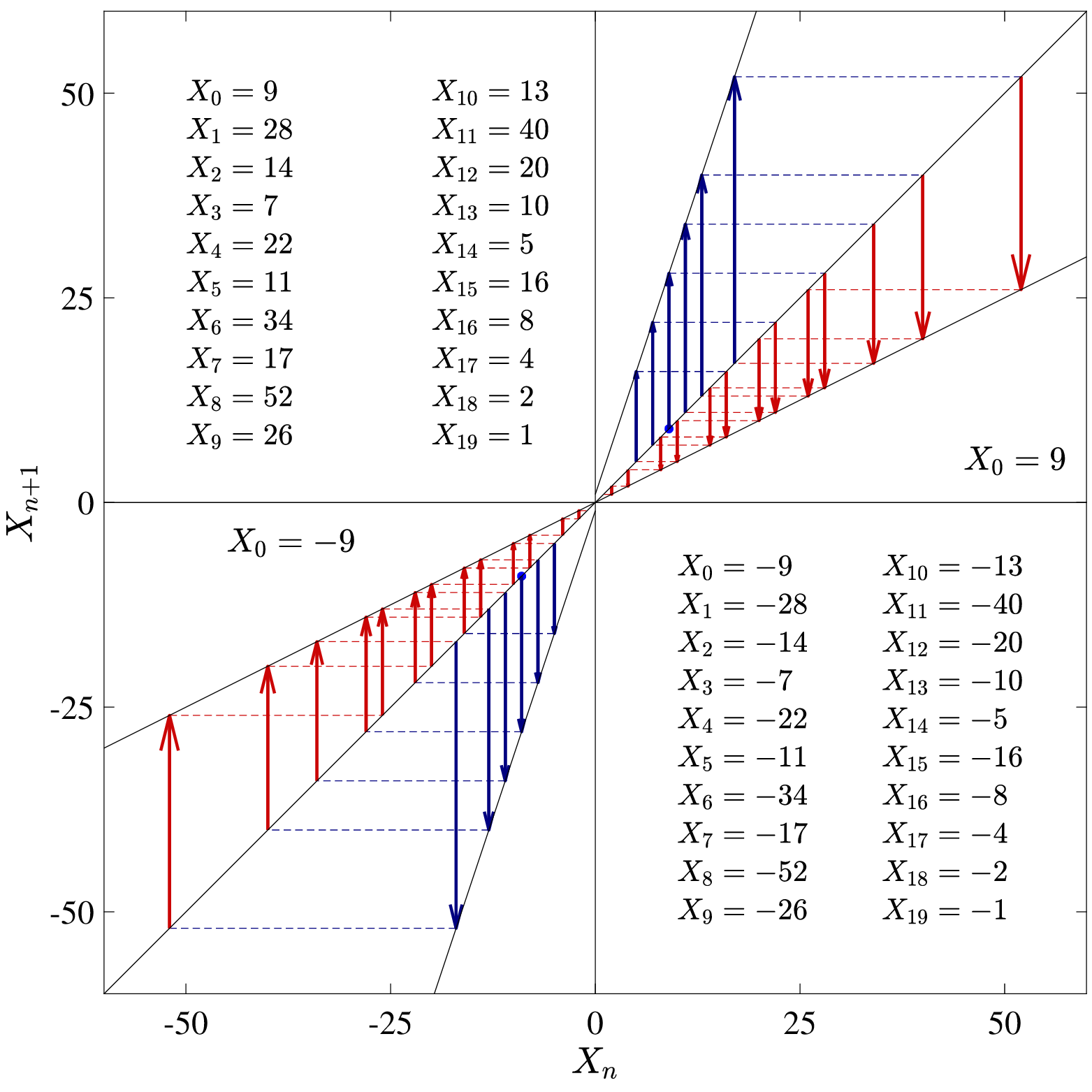}
  \caption{A cobweb plot for orbits of $X_0=\pm9$. In the second and fourth quadrants give the values of iterative sequences for $X_0=9$ and $X_0=-9$, respectively. }\label{fig-Map_PN}
\end{figure}

Figure~\ref{fig-Map_PN} illustrates symmetric trajectories for \( X_0 = \pm 9 \), converging to \(\pm 1\). This work focuses on \( X_0 \in \mathbb{Z}^+ \); subsequent sections detail the theoretical framework.

\section{Theoretical analysis}\label{sec:3}

We define two infinite-dimensional vectors:
\begin{equation}\label{eq-Bvector}
\boldsymbol{B} = [2^0, 2^1, 2^2, \dots, 2^m, \dots],
\end{equation}
\begin{equation}\label{eq-Ovector}
\boldsymbol{O} = [3, 5, 7, \dots, 2n+1, \dots],
\end{equation}
and the matrix $\boldsymbol{D} = \boldsymbol{B}^{\rm T} \boldsymbol{O}$, explicitly structured as
\begin{equation}\label{eq-Dmatrix}
\boldsymbol{D} =
\begin{bmatrix}
3 \cdot 2^0 & 5 \cdot 2^0 & \dots & (2n+1) \cdot 2^0 & \dots \\
3 \cdot 2^1 & 5 \cdot 2^1 & \dots & (2n+1) \cdot 2^1 & \dots \\
\vdots & \vdots & \ddots & \vdots & \vdots \\
3 \cdot 2^m & 5 \cdot 2^m & \dots & (2n+1) \cdot 2^m & \dots \\
\vdots & \vdots & \ddots & \vdots & \vdots
\end{bmatrix}.
\end{equation}

Let $\mathbb{B}$, $\mathbb{O}$, and $\mathbb{D}$ denote the sets formed by the elements of $\boldsymbol{B}$, $\boldsymbol{O}$, and $\boldsymbol{D}$, respectively. By Sharkovsky's ordering \cite{sharkowskii1964co}, $\mathbb{D} \cup \mathbb{B} \equiv \mathbb{Z}^+$.

It is clear that if $X_0 = 2^m \in \mathbb{B}$, then $M^m(X_0) = 1$ with $N_{\uparrow} = 0$ and $N = N_{\downarrow} = m$ (trivial trajectory); if $X_0 = (2n+1) \cdot 2^m \in \mathbb{D}$, then $M^m(X_0) = 2n+1 \in \mathbb{O}$. Thus, analyzing the Collatz map reduces to studying initial values $X_0 \in \mathbb{O}$.

\subsection{Estimation of iteration steps}

$\bullet$~\emph{Case I}: $N_{\uparrow} = 1$.

For $X_0 \in \mathbb{O} $ with $N_{\uparrow} = 1 $, we derive
\begin{equation}\label{eq-F1}
  X_0 = F_1(m) = \frac{2^m - 1}{3} \in \mathbb{O} \quad (m > 2).
\end{equation}
It is easily to prove that $F_1(m)\in \mathbb{O}$ iff $m = 2p $, yielding
\begin{equation}\label{eq-F1p}
F_1(p) = \frac{4^p - 1}{3}, \quad p \geq 2,
\end{equation}
with recursion
\begin{equation}
  F_1(p) = 4^{p-1} + F_1(p-1).
\end{equation}
The iteration steps for convergence is given by
\begin{equation}\label{eq-N1}
  N_1 =N_{\downarrow}+N_{\uparrow}=1+2p=1+\log_2(3X_0+1).
\end{equation}
Note that $F_1(1)\equiv 1\in\mathbb{B}$ from Eq.~(\ref{eq-F1p}).

$\bullet$~\emph{Case II}: $N_{\uparrow} = 2$.

For $X_0 \in \mathbb{O} $ with $N_{\uparrow} = 2 $, we have
\begin{equation}\label{eq-F2}
X_0 = F_2(p,k_1) = \frac{2^{k_1} F_1(p) - 1}{3}\quad (k_1\geq1),
\end{equation}
It is evident that
\begin{equation}\label{eq-F2pk}
F_2(p,k_1) = F_2(p,k_1 - 2) + 2^{k_1 - 2} F_1(p),
\end{equation}
and $k_1 = \log_2(3X_0 + 1) - \log_2(X_0)$. Thus, the iteration steps for convergence is
\begin{equation}\label{eq-N2}
N_2 = 2 + 2p + k_1 = 2 + 2\log_2(3X_0 + 1) - \log_2(X_0).
\end{equation}

$\bullet$~\emph{General Case}: $N_{\uparrow}=s$.

For $X_0 \in \mathbb{O} $ with $N_{\uparrow} = s $, recursively define
\begin{equation}\label{eq-Fs}
X_0 = F_s(p, k_{s-1}) = \frac{2^{k_{s-1}} F_{s-1}(p, k_{s-2}) - 1}{3},
\end{equation}
with $k_{s-1} = \log_2(3X_0 + 1) - \log_2(X_0) \in\mathbb{Z}^{+}$. The function $ F_s $ follows the recursion relation
\begin{equation}\label{eq-Fspk}
  F_s(p, k_{s-1}) = F_s(p, k_{s-1} - 2) + 2^{k_{s-1} - 2} F_{s-1}(p, k_{s-2}).
\end{equation}
Total iterations follow
\begin{equation}\label{eq-Ns}
\begin{aligned}
N_s &= N_{\uparrow} + N_{\downarrow} = s + 2p + \sum_{j=1}^{s-1} k_j \\
&= s + s \log_2(3X_0 + 1) - (s - 1) \log_2(X_0) \\
&= s \left[ 1 + \log_2\left( 3 + {1}/{X_0} \right) \right] + \log_2(X_0),
\end{aligned}
\end{equation}
implying logarithmic growth of $N_s$ with $X_0$.

Actually, the function $F_s$ can be expressed in an explicit form. For example, $F_4$ can be written as
\begin{equation}
\begin{aligned}
  &F_4(p, k_3) = \frac{2^{k_3} \left\{ \frac{2^{k_2} \left[ \frac{2^{k_1} \left( \frac{4^p - 1}{3} \right) - 1}{3} \right] - 1}{3} \right\} - 1}{3} = \\
  &\frac{2^{2p + k_1 + k_2 + k_3}}{3^4} - \frac{2^{k_1 + k_2 + k_3}}{3^4} - \frac{2^{k_2 + k_3}}{3^3} - \frac{2^{k_3}}{3^2} - \frac{1}{3}.
\end{aligned}
\end{equation}
The function $F_s$ can be further abbreviated as
\begin{equation}\label{eq-Fs-explicit}
  F_s = \boldsymbol{a}_s \cdot \boldsymbol{b}_s,
\end{equation}
where $\boldsymbol{a}_s \cdot \boldsymbol{b}_s $ denotes the inner product, and
\begin{equation}\label{eq-Fs-explicit-ab}
\begin{aligned}
  \boldsymbol{a}_s &= \frac{1}{3^s} [1, -3^0, -3^1, \dots, -3^{s-3}, -3^{s-2}, -3^{s-1}], \\
  \boldsymbol{b}_s &= [2^{k_s}, 2^{k_{s-1}}, 2^{k_{s-2}}, \dots, 2^{k_2}, 2^{k_1}, 2^0],
\end{aligned}
\end{equation}
with $ k_s > k_{s-1} > k_{s-2} > \dots > k_1 \geq 1 $. A similar function was presented in Ref. \cite{wirsching1998dynamical}. Note that if $ X_0 = F_s $, then $ N_{\downarrow} = k_s $, resulting in $ N_s = s + k_s $.

In fact, the behavior of $N_s/X_0$ is more important than the specific magnitude of $N_s$. For instance, for $ X_0 = 2^m \in \mathbb{B} $ and $ m \to \infty $, we have $ N_0 = m \to \infty $. However, we can also observe that
\begin{equation}\label{eq-N0}
  \lim_{m \to \infty} \frac{N_0}{X_0} = \lim_{m \to \infty} \frac{m}{2^m} = 0,
\end{equation}
implying \emph{finite} iterations for convergence.

For a give $s$, Eq.~(\ref{eq-Ns}) leads to
\begin{equation}\label{eq-Ns2X0}
  \lim_{X_0 \to \infty} \frac{N_s}{X_0} = \lim_{X_0 \to \infty} \frac{1}{X_0 \ln 2} = 0,
\end{equation}
validating convergence for $X_0 \in \mathbb{F}$, that is, the Collatz conjecture holds for $ X_0 \in \mathbb{F} $, where $\mathbb{F}$ is the set of all integers generated by the family of functions $F_s$. Equation~(\ref{eq-N0}) and (\ref{eq-Ns2X0}) provide \emph{proof of convergence}, aligning with Tao's probabilistic result \cite{tao2022orbitscollatzmapattain}.

Above analysis shows that if $\mathbb{F} \equiv \mathbb{O} \cup \{1\}$, that is, \emph{if all odd numbers can be generated by $F_s$, the Collatz conjecture holds universally}. Proving this equivalence would indirectly establish uniqueness of the cycle $\{4,2,1\}$. However, providing a direct proof of this point is difficult, and efforts will continue in the future.

\subsection{Definition of attraction basins}

The function $F_s$ exhibits self-similarity: for any $s$, iterations eventually enter the decaying orbit $X_{N_s - 2p} = 4^p \in \mathbb{B}$. Specifically, $X_{N_s - 2p - 1} = \frac{4^p - 1}{3} \in \mathbb{O}$ acts as the entry point to the period-three cycle. For $X_0 \in \mathbb{O}$, $X^p = F_1(p) = \frac{4^p - 1}{3}$ functions as a black hole---once encountered, the map irreversibly transitions to $\mathbb{B}$.

We \emph{define} the set of $\{X_0\}$ sharing the same $p$-value as the basin of attraction for $p$, with $X^p$ termed an \emph{attractor}. Below, we analyze basin properties via $F_s$.

Non-trivial case ($p \geq 2$):
For $F_2(p, k_1)$, initial values $F_2(p,1)$ and $F_2(p,2)$ determine all $F_2(p,k_1)$ via the recursion relation (\ref{eq-F2pk}).
Specific solutions include:
\begin{equation}\label{eq-F2ss}
\begin{aligned}
F_2(1,1) &= \frac{1}{3}, ~ &F_2(2,1) &= 3, ~ &F_2(3,1) &= \frac{41}{3}, \\
F_2(1,2) &= 1, ~ &F_2(2,2) &= \frac{19}{3}, ~ &F_2(3,2) &= \frac{83}{3}.
\end{aligned}
\end{equation}
The recursion further satisfies
\begin{equation}\label{eq-F2p2k1}
\begin{cases}
F_2(p+3,1) = 7 \cdot 2^{2p+1} + F_2(p,1), \\
F_2(p+3,2) = 7 \cdot 2^{2p+2} + F_2(p,2).
\end{cases}
\end{equation}

Equations~(\ref{eq-F2ss}) and (\ref{eq-F2p2k1}) provide key observations:

(i). $F_2(p,k_1) \in \mathbb{O}$ iff: $p = 2 + 3m$ and $k_1 = 2n - 1$, or $p = 4 + 3m$ and $k_1 = 2n$, where $m \in \mathbb{Z}$, $n \in \mathbb{Z}^+$.

(ii). For $p = 3n$, $F_2(p,k_1)$ yields fractions, implying $X_0 = X^p$ uniquely resides in the basin of $p = 3n$.

\subsection{The scheme of like-Bernoulli shift map}

\begin{figure}[t]
  \centering
  \includegraphics[width=1\columnwidth]{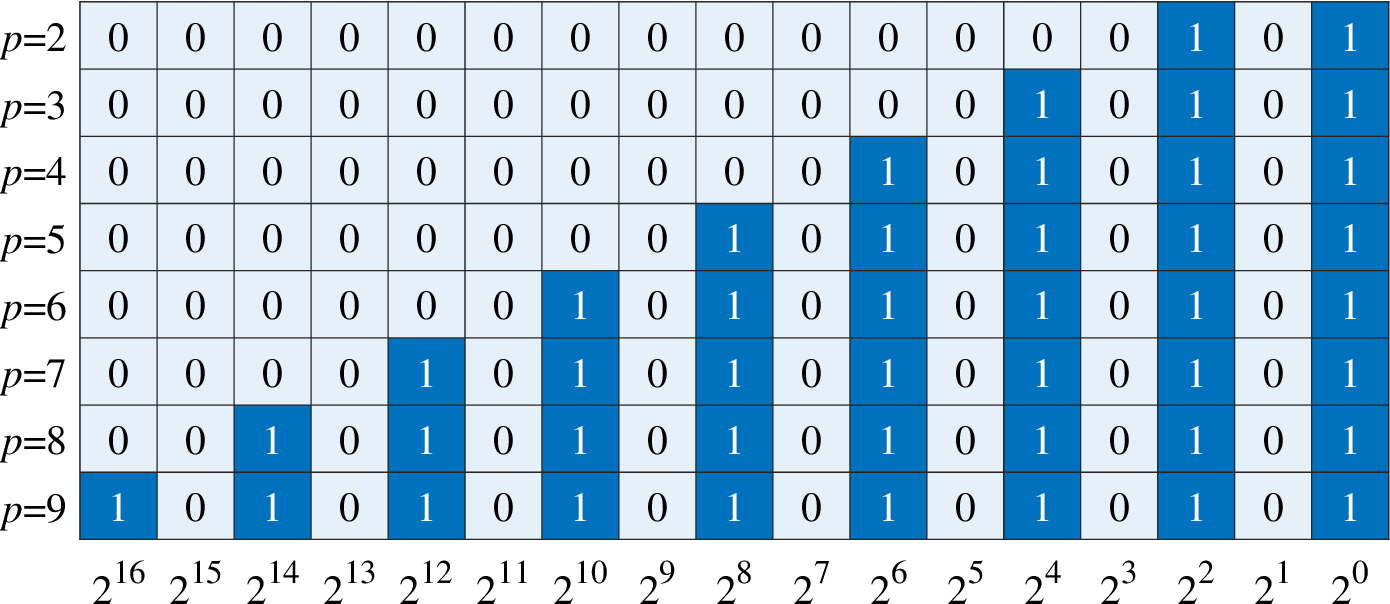}
  \caption{The binary representations of $X^p=(4^p-1)/3=$5, 21, 85, 341, 1365, 5461, 21845, 87381 for $p=2,3,\dots,9$. }\label{fig-heatmap}
\end{figure}

\begin{figure}[t]
  \centering
\includegraphics[width=1\columnwidth]{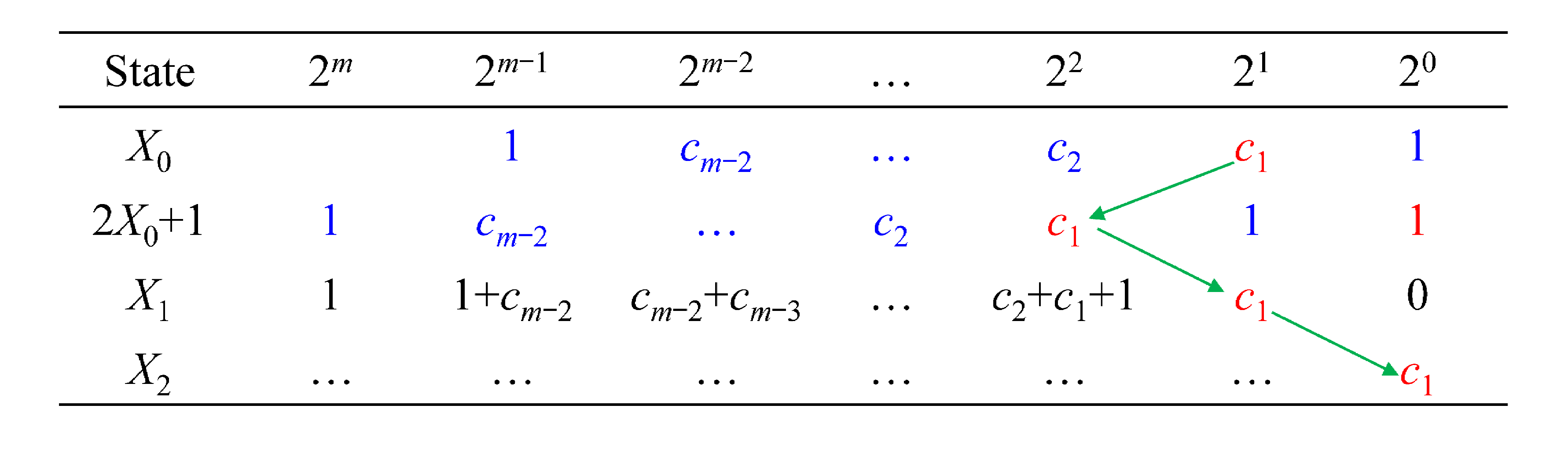}
  \caption{Demonstration of iterations in the binary system resembling a Bernoulli shift scheme, where $X_2 = M(X_1) = M^2(X_0)$, illustrating how the operation $M^2$ moves $c_1$ to the rightmost end.}\label{fig_tab1}
\end{figure}

Let $X_0\in \mathbb{O}$. The Collatz map yields
\begin{equation*}
\begin{cases}
X_1 = 3X_0 + 1, \\
X_2 = {X_1}/{2} = M^2(X_0),
\end{cases}
\end{equation*}
where $X_1$ is necessarily even (see Fig.~\ref{fig_tab1}). For $2^{m-1} < X_0 < 2^m$, express $X_0$ in binary as
\begin{equation*}
X_0 = 2^{m-1} + \sum_{j=1}^{m-2} c_j 2^j + 2^0 \quad (c_j \in \{0,1\}, \, c_0 = c_{m-1} = 1).
\end{equation*}
Figure~\ref{fig-heatmap} reveals self-similarity binary patterns in attractors $X^p$ for $p = 2, 3, \dots, 9$.

From Fig.~\ref{fig_tab1}, we see that $2X_0 + 1$ corresponds to a leftward binary shift with a terminal 1 appended; and $X_1 = X_0 + (2X_0 + 1)$, inducing binary coefficient addition, where carry operations occur when $c_j \geq 2$, propagating to higher bits, e.g., $c'_2 = (c_2 + c_1 + 1) \mod 2$, $c'_3 = c_3 + \lfloor (c_2 + c_1 + 1)/2 \rfloor$. We clearly observe a shift dynamics: $M^2$ shifts $c_1$ of $X_0$ to the position of $c_0$ of $X_2$, updating other coefficients via addition, and $N_{\uparrow}$ increments by 1 and repeats the shift operation for $X_2$ similarly to that for $X_0$, i.e., $M^2$ followed by $M^2$ if $c_1 = 1$; otherwise, $M^2$ followed by $M$. Statistically, a rightward shift occurs due to the equiprobability of $c_j \in \{0,1\}$---validating the Collatz conjecture. Additionally, the ergodicity of the binary shift map \cite{book2011Ergodic} ensures that the state of $X^p$ (Fig.~\ref{fig-heatmap}) must eventually appear, further guaranteeing the validity of the Collatz conjecture.

\section{Numerical verifications}\label{sec:4}

Based on the theoretical analysis above, we only need to consider the case where $X_0\in\mathbb{O}$
in the numerical simulations. Since the set $\mathbb{O}$
contains an infinite number of elements, in practical purposes, we restrict $X_0$
to finite intervals in the simulations and conduct statistical analysis. For example, we choose
$X_0\in\{3,5,\dots,2L+1\}$ and then count how many numbers belong to the basin of attraction of $p$, where $L$ is a positive integer that controls the size of the statistics window.

\begin{figure}[t]
  \centering
  \includegraphics[width=1\columnwidth]{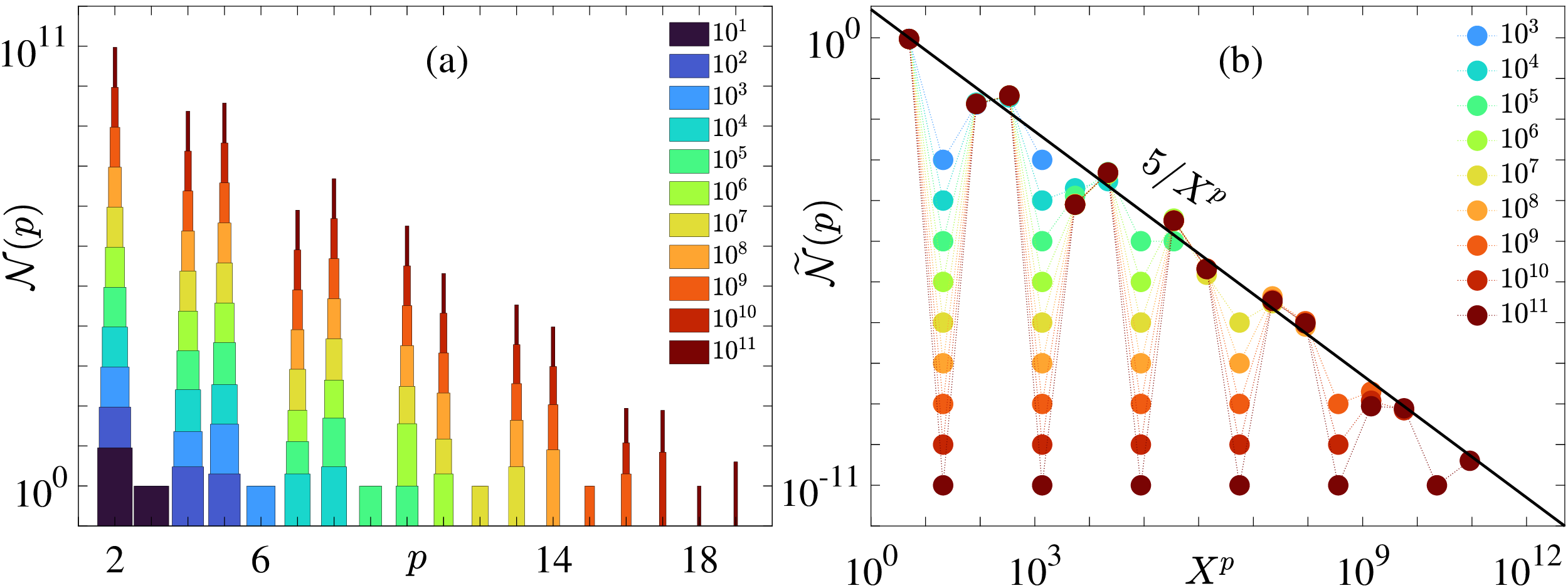}
  \caption{(a) Histograms of number $\mathcal{N}(p)$ in different attraction basins $p$ within a given range of $X_0\in\{3,5,\dots,2L+1\}$. The legend shows the value of $L$. (b) Same as the panel (a) but in normalized form $\widetilde{\mathcal{N}}(p)=\mathcal{N}(p)/L$ and the horizontal axis is replaced by $X^p=F_1(p)$. The solid line serves as reference.}\label{fig_doorCount}
\end{figure}

Figure \ref{fig_doorCount}(a) shows the count $\mathcal{N}(p)$ as a function of $p$ for various values of $L$. We observe that, regardless of the value of $L$, $\mathcal{N}(2)$ is always the largest. In Addition, we see that $\mathcal{N}(p)=1$ when $\text{mod}(p, 3) =0$, which is consistent with the theoretical analysis presented in subsection B of Sec.~\ref{sec:3}.
Figure \ref{fig_doorCount}(b) plots the normalized count $\widetilde{\mathcal{N}} = \mathcal{N}/L$, representing the percentage of numbers in each basin of attraction. Notice that, except for cases where $\text{mod}(p, 3)=0$, $\widetilde{\mathcal{N}}(p)\propto 1/X^p$, refer to the black reference line. For larger values of $L$, $\widetilde{\mathcal{N}}(2) \approx 93.77\%$, as detailed in Tab.~\ref{tab2}.

\begin{table*}[t]
\centering
\caption{Statistics of numbers in different attraction basins $p$ within a given range of $X_0\in[3,5,\dots,2L+1]$.}\label{tab2}
\begin{tabular}{lrrrrrrrrrrr}
\toprule
\toprule
$p$ & $L=10^1$&$10^2$&$10^3$&$10^4$&
$10^5$&$10^6$&$10^7$&$10^8$&$10^9$&$10^{10}$&$10^{11}$\\
\midrule
2&~~9&~~94&~~940&~~9395&~~93679&~~938003&~~9378361&~~93772537&~~937676531&~~9377184597&~~93774780663\\
\underline{\emph{3}}&1&1&1&1&1&1&1&1&1&1&1\\
4&0&3&23&255&2412&23743&237828&2373777&23761165&237454856&2373306190\\
5&0&2&35&343&3842&37687&377838&3793838&37961580&379374692&3792124780\\
\underline{\emph{6}}&0&0&1&1&1&1&1&1&1&1&1\\
7&0&0&0&2&13&78&830&8098&80062&796409&7937609\\
8&0&0&0&3&50&448&4810&48229&485728&4843192&48398337\\
\underline{\emph{9}}&0&0&0&0&1&1&1&1&1&1&1\\
10&0&0&0&0&1&36&311&3253&32329&320937&3199254\\
11&0&0&0&0&0&2&15&212&2127&20883&206700\\
\underline{\emph{12}}&0&0&0&0&0&0&1&1&1&1&1\\
13&0&0&0&0&0&0&3&44&363&3419&34128\\
14&0&0&0&0&0&0&0&8&108&974&9551\\
\underline{\emph{15}}&0&0&0&0&0&0&0&0&1&1&1\\
16&0&0&0&0&0&0&0&0&2&12&88\\
17&0&0&0&0&0&0&0&0&0&7&78\\
\underline{\emph{18}}&0&0&0&0&0&0&0&0&0&0&1\\
19&0&0&0&0&0&0&0&0&0&0&4\\
\bottomrule
\bottomrule
\end{tabular}
\end{table*}

In principle, the value of $p$ can extend to infinity, so it is reasonable to infer that the numerical distribution in Tab.~\ref{tab2} will resemble the form of an upper triangular matrix. Of course, to observe non-zero values for larger $p$, a larger $L$ is required. Based on the data, it is conjectured that as $p \to \infty$ and $L\to \infty$, then $\widetilde{\mathcal{N}}(2) \approx 93.77\%$, $\widetilde{\mathcal{N}}(4) \approx 2.37\%$, and $\widetilde{\mathcal{N}}(5) \approx 3.79\%$, that is, $\widetilde{\mathcal{N}}(2)+\widetilde{\mathcal{N}}(4)+\widetilde{\mathcal{N}}(5) \approx 99.93\%$ remains essentially unchanged, and increasing $p$ and $L$ only affects the significant digits of the decimal point.
It is clear that \emph{the size of basin of attraction follows a power-law distribution}.

\begin{figure*}[t]
  \centering
  \includegraphics[width=2\columnwidth]{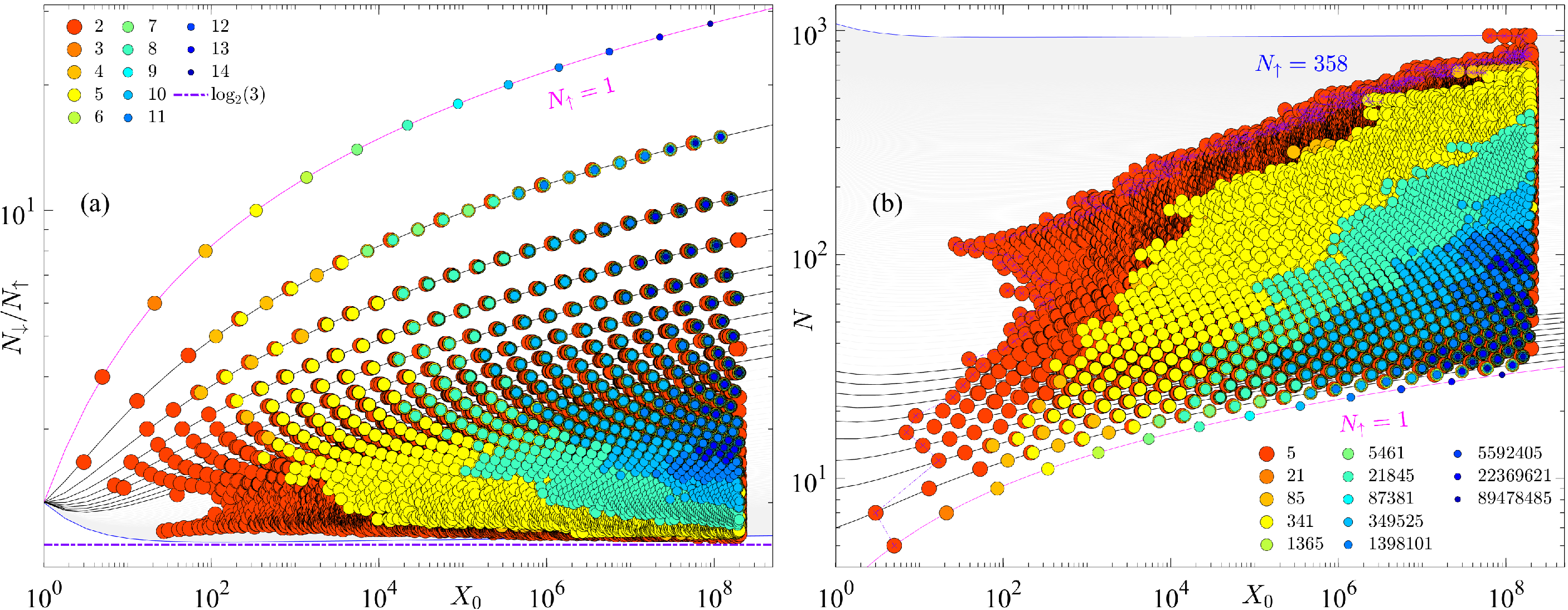}\\
  \caption{The dependence of $N_{\downarrow}/N_{\uparrow}$ (a) and $N=N_{\downarrow}+N_{\uparrow}$ (b) on $X_0=2n+1$ for $n=1,2,\dots,10^8$. Different colored points indicate belonging to different attraction basins $p$, see the legend in panel (a), while the legend in panel (b) shows the value of $F_1(p)$. The solid lines are the theoretical predictions given by expression (\ref{eq-Ns}). The dotted purple line marked by cross in panel (b) gives the upper boundary of the region, which is replotted in Fig.~\ref{fig-edge}(a).}\label{fig-Nsum}
\end{figure*}

Figure~\ref{fig-Nsum}(a) diaplays the ratio $N_{\downarrow}/N_{\uparrow}$ as a function of $X_0$. Different colored points represent numbers belonging to different basins of attraction. The solid lines correspond to theoretical predictions given by Eq. (\ref{eq-Ns}). It is seen that the theoretical and numerical results align very well. Additionally, the ratio approaches $\log_2(3) \approx 1.585$ as $N_{\uparrow}$ increases, as indicated by the purple horizontal dashed line. Figure \ref{fig-Nsum}(b) shows the dependence of the total number of iterations, $N$, on $X_0$. Note that $N$ grows slowly as $X_0$ increases. The numerical results are fully consistent with the theoretical predictions, as shown by the solid lines from Eq.~(\ref{eq-Ns}). This supports the conclusion that the number of iterations required for convergence is indeed a finite value relative to $X_0$, confirming Eqs.~(\ref{eq-N0}) and (\ref{eq-Ns2X0}). Furthermore, in Fig.~\ref{fig-Nsum}(b), the purple crossed-dotted line marks the upper boundary of the region, which consists of the points with the smallest $X_0$ for a given $N_{\uparrow}$. We observe that the boundaries of the basins of attraction for different $p$ resemble coastlines and exhibit a high degree of similarity. Notably, the distribution of the basin of attraction for $p = 2$ nearly covers the entire range of the calculated interval, which is consistent with the statistical results presented in Fig.~\ref{fig_doorCount}. Specifically, the numbers on the upper boundary all belong to the basin of attraction of $p = 2$.

\begin{figure}[t]
  \centering
  \includegraphics[width=1\columnwidth]{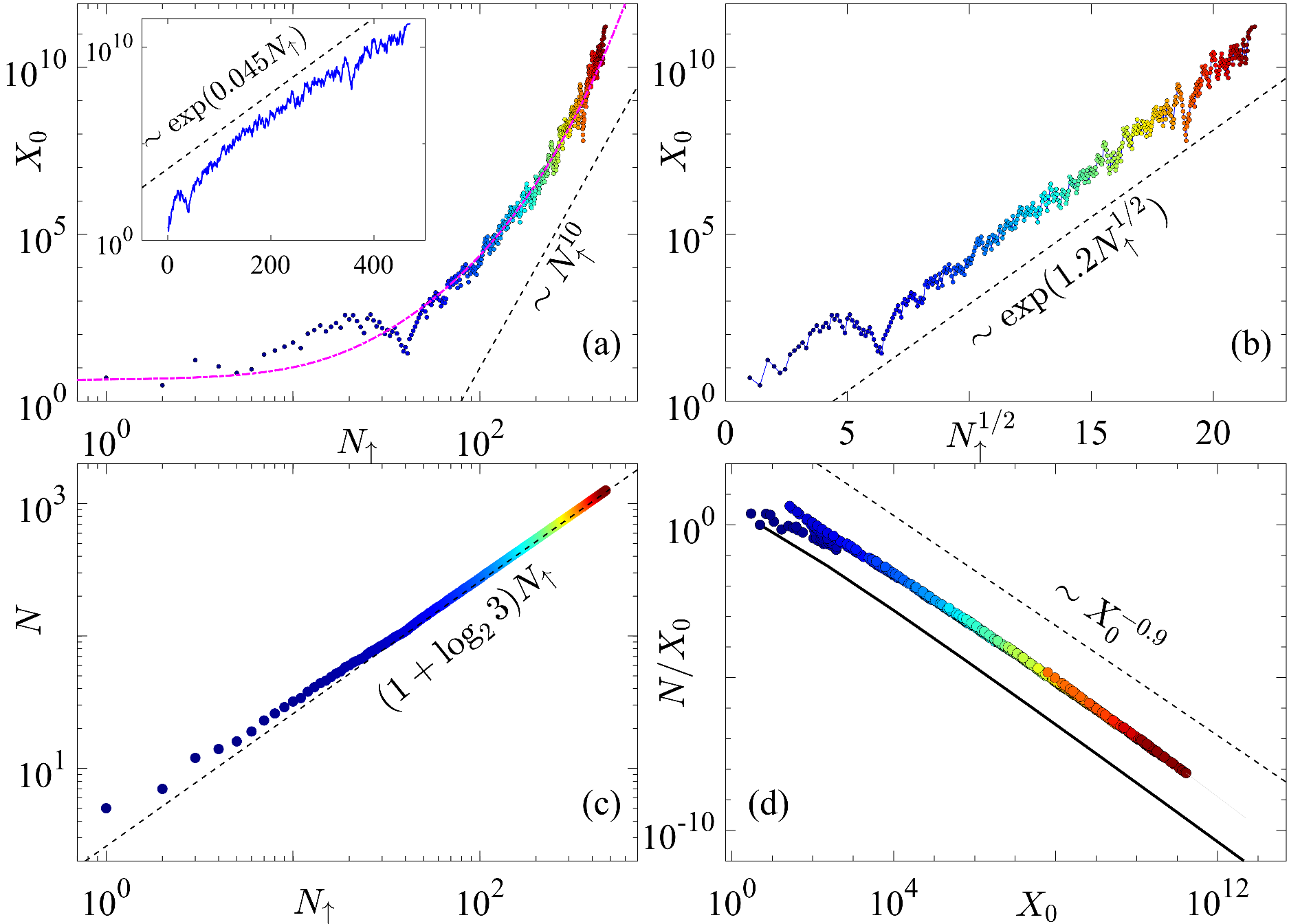}\\
  \caption{(a) $X_0$ of the upper boundary points in Fig.~\ref{fig-Nsum}(b) as a function of $N_{\uparrow}$ in log-log scale. The magenta dashed line is a plot of a stretched exponential function $\exp(1.2N_{\uparrow}^{1/2})/7+4$ for reference. Inset: Same as the main panel but in semi-log scale. (b) Same as the panel (a) but the abscissa is rescaled as $N_{\uparrow}^{1/2}$.
  The data in panels (c) and (d) are the same as that in panel (a) but shows in different forms. (c) $N=N_{\uparrow}+N_{\downarrow}$ as a function of $N_{\uparrow}$. (d) The dependence of $N_{\uparrow}/X_0$ on $X_0$. The black line is given by formula (\ref{eq-Ns}) for $N_{\uparrow}=1$. The dashed lines in all panels are plotted for reference.}\label{fig-edge}
\end{figure}

To investigate the properties of the numbers at the upper boundary marked in Fig.~\ref{fig-Nsum}(b) in more detail, we present the behavior of these numbers over a broader range in Fig. \ref{fig-edge}. The abscissa ($X_0$) and ordinate ($N$) of the boundary points in Fig.~\ref{fig-Nsum}(b) as functions of $N_{\uparrow}$ are shown in Figs.~\ref{fig-edge}(a)-(c). Figure~\ref{fig-edge}(a) demonstrates that $X_0$ increases nearly in a stretched exponential manner with $N_{\uparrow}$ (see the magenta dashed curve), which is faster than a power-law growth (black dashed line), but slower than exponential growth (black dashed line in the inset). Figure~\ref{fig-edge}(b) provides a clearer depiction of how $X_0$ increases with $N_{\uparrow}$, i.e., $X_0\propto \exp(1.2N_{\uparrow}^{1/2})$. Additionally, a self-similarity fractal structure, reminiscent of a coastline, is observed in the distribution of $X_0$.

Figure~\ref{fig-edge}(c) shows the dependence of $N$ on $N_{\uparrow}$, where $N$ increases approximately linearly with $N_{\uparrow}$. From Eq.~(\ref{eq-Ns}), for a large $X_0$, we have $N \approx [1 + \log_2(3)] N_{\uparrow}$, which agrees with the numerical results very well. Figure~\ref{fig-edge}(d) displays the ratio $N/X_0$ as a function of $X_0$, where the black line represents the theoretical prediction derived from Eq.~(\ref{eq-Ns}). It is clear that as $X_0 \to \infty$, $N/X_0 \to 0$, indicating that the number of iterations required to converge to the period-three orbit becomes negligible relative to large $X_0$. In this sense, the Collatz conjecture holds for all positive integers.

\begin{figure}[t]
  \centering
  \includegraphics[width=1\columnwidth]{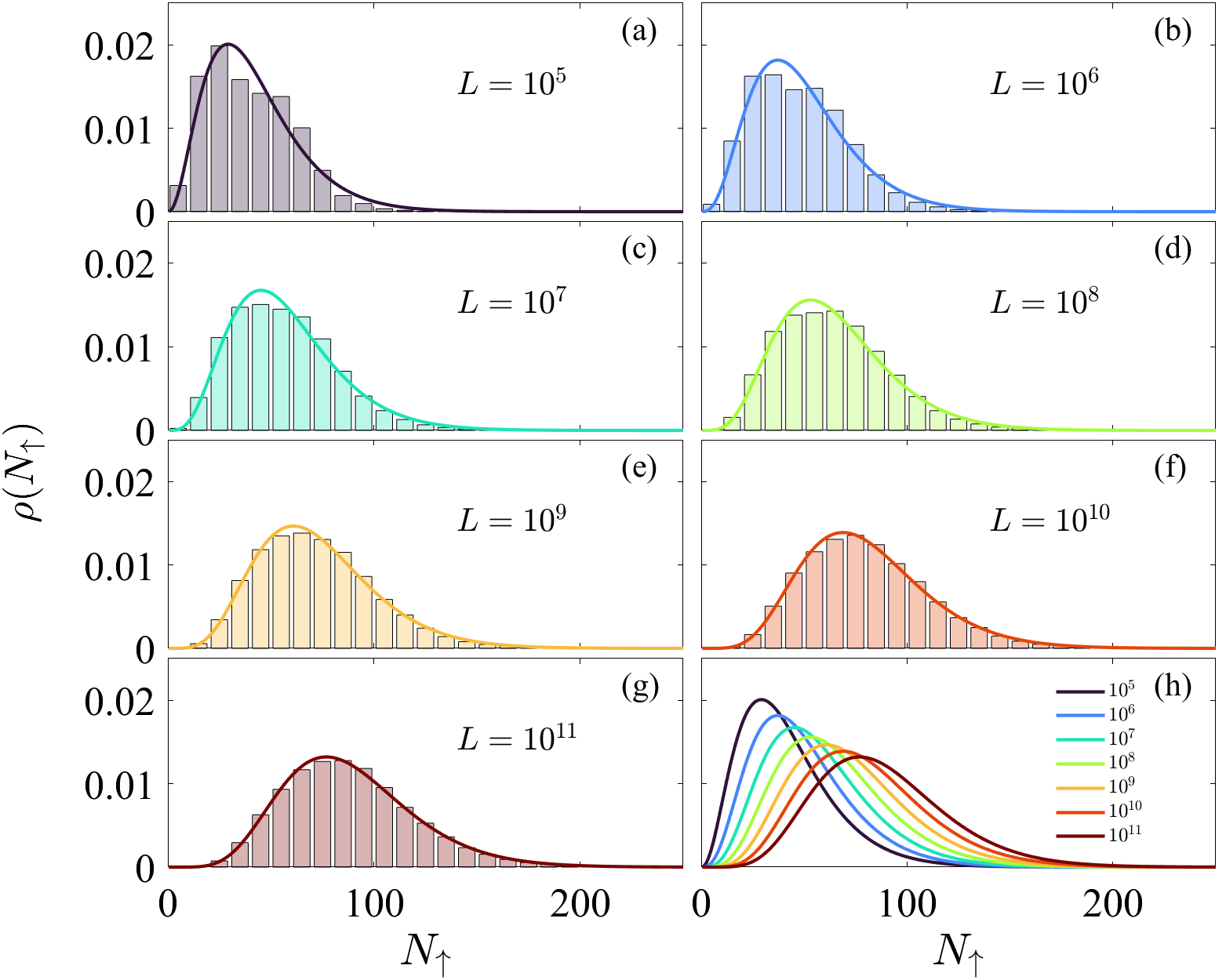}
  \caption{(a)-(g) The distribution of numbers over $N_{\uparrow}$ within a given interval of $X_0\in[3,5,\dots,2L+1]$. The solid line is a fitting curve based on the Gamma distribution with the form of $\rho(x)=\frac{A}{\Gamma(K)\theta^K}x^{K-1}e^{-x/\theta}$. (h) Summarize all fitted curves together for comparison.}\label{fig-distribution}
\end{figure}

Figure~\ref{fig-distribution} shows the distribution of numbers over $N_{\uparrow}$ for a given range of $X_0 \in [3, 5, \dots, 2L + 1]$. In other words, the height of the histogram represents the proportion of numbers with the same $N_{\uparrow}$. From Figs.~\ref{fig-distribution}(a)-(g), we observe that the distribution approximately follows a Gamma distribution with the form $\rho(x) = \frac{A}{\theta^K \Gamma(K)} x^{K-1} e^{-x/\theta}$ (see the solid line in each panel). Moreover, the distribution function becomes flatter as $L$ increases, as shown by the solid lines in Fig.~\ref{fig-distribution}(h).

\begin{figure}[t]
  \centering
  \includegraphics[width=1\columnwidth]{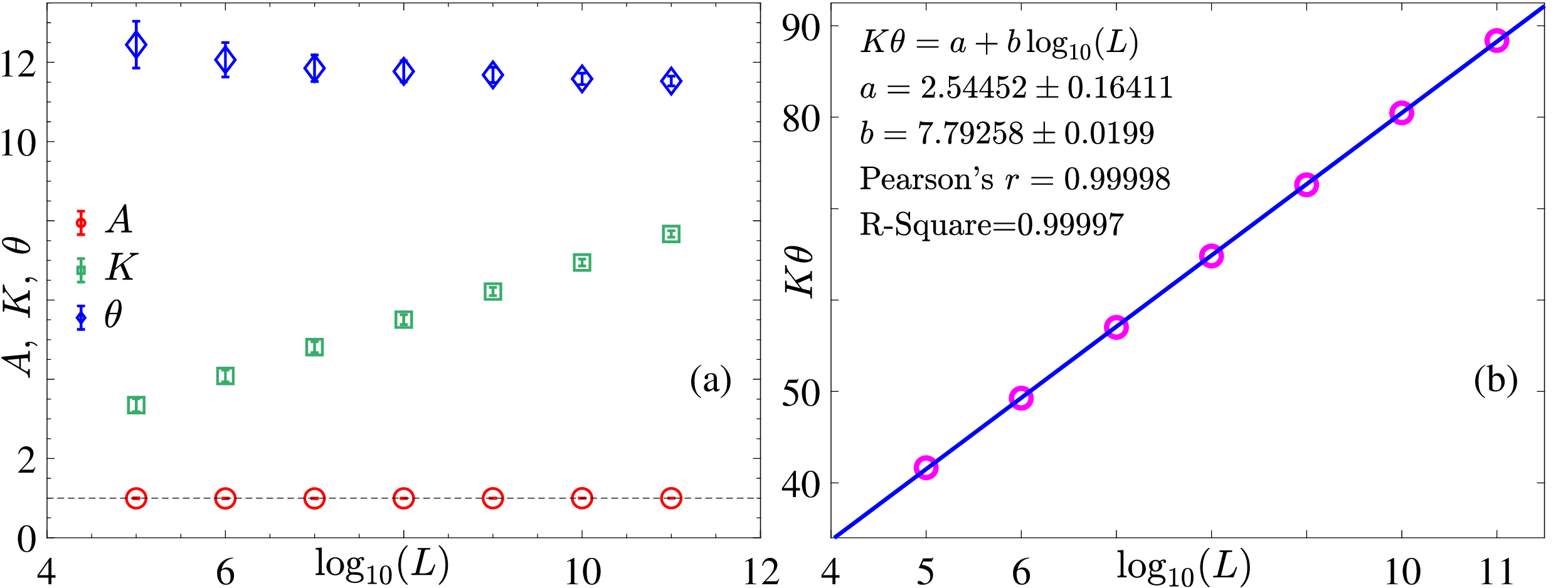}
  \caption{(a) The fitting parameters of the Gamma distribution function in Fig.~\ref{fig-distribution} as a function of $L$. (b) Dependence of the mean value of the Gamma distribution $K\theta$ on $L$.}\label{fig-gamma_AKtheta}
\end{figure}

Figure \ref{fig-gamma_AKtheta}(a) shows the dependence of the parameters of the Gamma distribution function on $L$. We observe that $A \simeq 1$ remains constant; $K$ increases as $L$ increases, concretely, $K \propto \log_{10}(L)$; and $\theta$ exhibits a slow decay, remaining approximately constant within the range studied. As is known, the mean value of the Gamma distribution is given by $K\theta$, which also increases as $L$ increases, as shown in Fig.~\ref{fig-gamma_AKtheta}(b). The best linear fitting suggests that the mean value of $\overline{N}_{\uparrow} = K\theta \approx 2.54 + 7.79 \log{10}(L)$.

\begin{figure}[t]
  \centering
  \includegraphics[width=1\columnwidth]{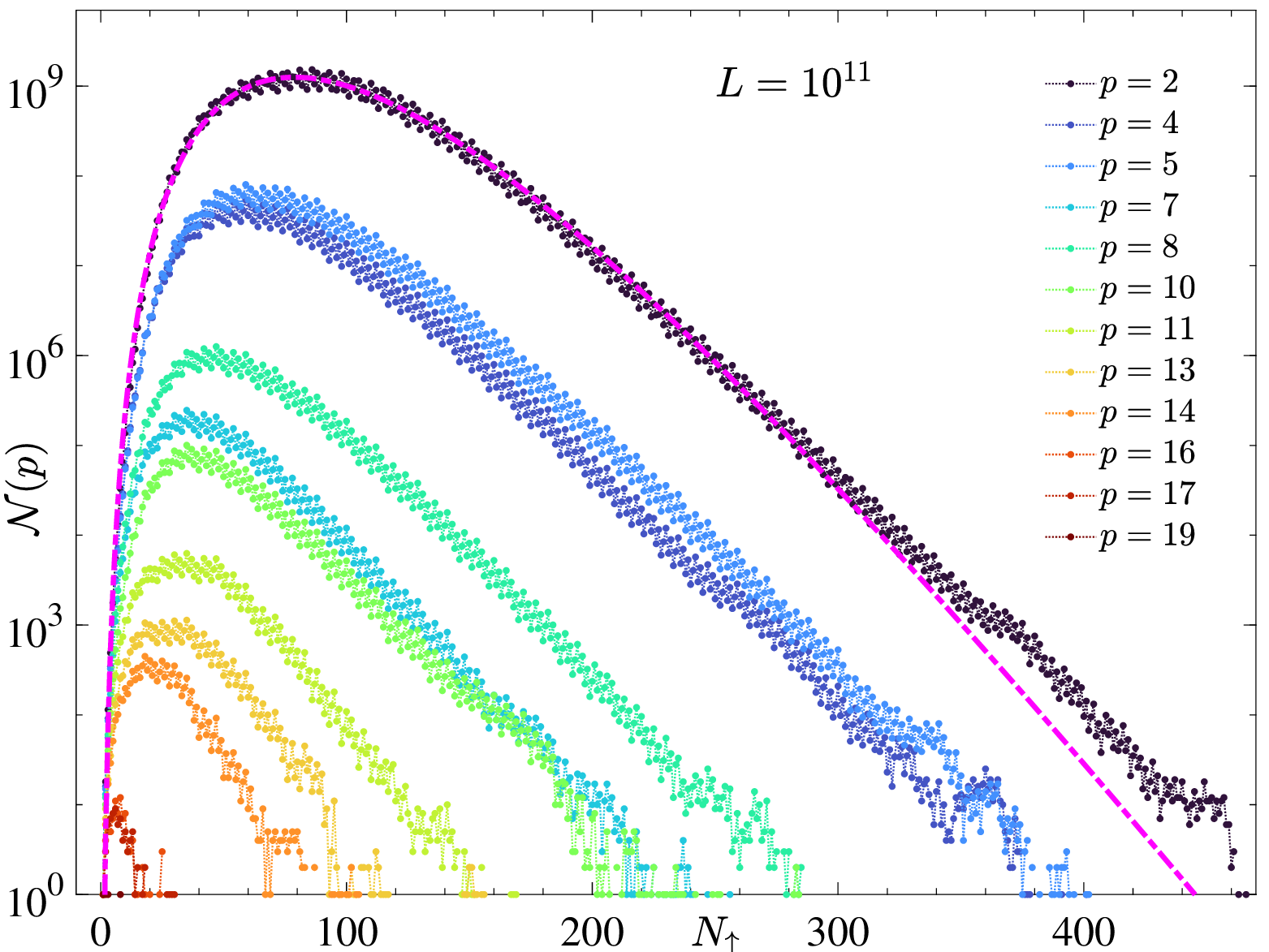}
  \caption{The count number $\mathcal{N}(p)$ over $N_{\uparrow}$ within a given interval of $X_0\in[3,5,\dots,2L+1]$ in a semi-log scale. The magenta dashed line is a fitting curve based on the Gamma distribution.}\label{fig-distribution_diffP}
\end{figure}

We further count the distribution of numbers in different basins of attraction over $N_{\uparrow}$ within the range $X_0 \in [3, 5, \dots, 2L+1]$, where $L= 10^{11}$ is fixed (we have verified that changing the value of $L$ produces qualitatively identical results). The numerical results are shown in Fig.~\ref{fig-distribution_diffP}, where $\mathcal{N}(p)$ represents the number of numbers belonging to the basin of attraction of $p$. It is evident that each $\mathcal{N}(p)$ follows a Gamma distribution, with different parameters for each $p$ (see the magenta dashed curve, which nearly covers all points for $p=2$).

\section{Summary and discussions}\label{sec:5}

In summary, we have investigates the Collatz map from the perspective of nonlinear dynamics, yielding four principal advances: First, by organizing positive integers via Sharkovsky's ordering, we demonstrate that analyzing odd initial values alone suffices to characterize Collatz dynamics. Second, we establish a classification system using ``direction phases'' to distinguish ascending and descending iteration patterns. This framework reveals a logarithmic scaling law between iteration steps and initial values through derived recursive functions $F_s$ (parameterized by upward phase count $s$), proving finite-time convergence to the period-three cycle. Third, we identify dynamical equivalence between the Collatz map and a binary shift map, whose ergodicity guarantees universal convergence to attractors. Fourth, extensive numerical simulations show power-law distributed attraction basins and Gamma-distributed odd numbers across upward phases. These results offer valuable insights into the dynamics of discrete systems and their connections to number theory.

Furthermore, our analysis indicates that proving all odd numbers ($>1$) are generated by the $F_s$ function family would confirm the Collatz conjecture as a theorem. Within our computational range, all tested odd numbers align with $F_s$ generation, supporting the conjecture that every odd number can be expressed in $F_s$'s form. However, a general proof of $F_s$'s universal generative capacity remains challenging and requires further investigation. Notably, studying $F_s$'s properties may provide novel perspectives on prime number distributions, given that all primes (except 2) belong to the odd number set theoretically covered by $F_s$.

\begin{acknowledgments}
This work was supported by the National Science Foundation of China (Grants No.~12465010, No.~12247106, No.~12005156, No.~11975190, and No.~12247101). W. Fu also acknowledges support from the Youth Talent (Team) Project of Gansu Province, the Long-yuan Youth Talents Project of Gansu Province, the Fei-tian Scholars Project of Gansu Province, the Leading Talent Project of Tianshui City, the Innovation Fund from the Department of Education of Gansu Province (Grant No.~2023A-106), and the Open Project Program of Key Laboratory of Atomic and Molecular Physics $\&$ Functional Material of Gansu Province (6016-202404).
\end{acknowledgments}

\bibliography{ref3xp1}

\end{document}